\documentclass[aps,twocolumn,nofootinbib,showpacs,hyperref]{revtex4-1}

\usepackage{amsfonts}
\usepackage{amsmath}
\usepackage{amssymb}
\usepackage{bm}
\usepackage{dcolumn}
\usepackage{graphicx}
\usepackage{graphics}
\usepackage[latin1]{inputenc}
\usepackage{latexsym}
\usepackage{rotating}
\usepackage{hyperref}
\usepackage[all]{hypcap} 
\usepackage{xspace} 
\usepackage[usenames]{color}
\usepackage{mathrsfs}

\usepackage{ulem}
\definecolor {darkgreen}{rgb}{0.2,0.7,0.2}

\newcommand\be{\begin{equation}}
\newcommand\ba{\begin{eqnarray}}
\newcommand\ee{\end{equation}}
\newcommand\ea{\end{eqnarray}}
\newcommand\bw{\begin{widetext}}
\newcommand\ew{\end{widetext}}

\newcommand{\nn}{\nonumber}

\newcommand{\pd}{\partial}

\newcommand{\nr}{{\mbox{\tiny nr}}}

\begin{document}
\title{Toward realistic and practical no-hair relations \\ for neutron stars in the nonrelativistic limit} 

\author{Katerina Chatziioannou}
\affiliation{Department of Physics, Montana State University, Bozeman, Montana 59717, USA.}

\author{Kent Yagi}
\affiliation{Department of Physics, Montana State University, Bozeman, Montana 59717, USA.}

\author{Nicol\'as Yunes}
\affiliation{Department of Physics, Montana State University, Bozeman, Montana 59717, USA.}

\begin{abstract}

The gravitational properties of astrophysical objects depend sensitively on their internal structure. In Newtonian theory, the gravitational potential of a rotating star can be fully described by an infinite number of multipole moments of its mass distribution. Recently, this infinite number of moments for uniformly-rotating stars were shown semianalytically to be expressible in terms of just the first three: the mass, the spin, and the quadrupole moment of the star. The relations between the various lower multipole moments were additionally shown to depend weakly on the equation of state, when considering neutron stars and assuming single polytropic equations of state. Here we extend this result in two ways. First, we show that the universality also holds for realistic equations of state, thus relaxing the need to use single polytropes. Second, we derive purely analytical universal relations by perturbing the equations of structure about an $n=0$ polytrope that reproduce semianalytic results to $\mathcal{O}(1\%)$. We also find that the linear-order perturbation vanishes in some cases, which provides further evidence and a deeper understanding of the universality.

\end{abstract}

\date{\today}
\maketitle

\section{Introduction}

Neutron stars (NSs) are among the most interesting astrophysical objects in Nature. The extremely high matter densities encountered in their interiors make them an ideal laboratory for extreme physics~\cite{lattimer-prakash-review}. This physics depends sensitively on the, yet unknown, NS equation of state (EoS), a relation between the pressure and the mass density\footnote{In general, the NS EoS depends on both the mass density and the temperature. However, here we are interested in NS binaries that are about to coalesce. These are old and cold (i.e. the temperature is much smaller than the Fermi temperature), and hence, it is sufficient to treat the EoS as barotropic, (i.e. the pressure only depends on the mass density). See~\cite{Martinon:2014uua} for a more generic study of proto-NSs with non-barotropic EoSs.}. The NS EoS has already been weakly constrained from type-I X-ray bursters with photospheric radius expansion, thermal spectra from transient low-mass X-ray binaries and nuclear physics experiments~\cite{steiner-lattimer-brown,Lattimer:2012nd,Lattimer:2013hma}. Many future observations, including X-ray flux detection emitted from hot spots on the NS surface~\cite{Psaltis:2013fha,Lo:2013ava} with NICER~\cite{2012SPIE.8443E..13G} and LOFT~\cite{2012SPIE.8443E..2DF} and gravitational wave measurements of NS binaries~\cite{flanagan-hinderer-love,read-love,hinderer-lackey-lang-read,lackey,damour-nagar-villain,Lackey:2013axa,Read:2013zra,favata-sys,kent-sys,DelPozzo:2013ala,Wade:2014vqa} with second-generation ground-based interferometers~\cite{ligo,virgo,kagra}, aspire to infer the NS EoS even more accurately and provide invaluable information about physics in extreme densities.

Recently, several approximately universal (i.e. approximately EoS-independent) relations were discovered between various EoS-dependent quantities that characterize NSs. Relations between the moment of inertia, the quadrupole moment and the tidal deformability (tidal Love number) were first discovered in~\cite{I-Love-Q-Science,I-Love-Q-PRD}; other approximately universal relations can be found in~\cite{andersson-kokkotas-1998,kokkotas-living,tsui-leung,lau,1994ApJ...424..846R,lattimer_prakash2001,bejger,lattimer-schutz,urbanec,1989ApJ...340..426L,1997PhR...280....1P,2004Sci...304..536L,2009A&A...502..605H,kiuchi,kyutoku,Bauswein:2011tp,baubock, Yagi:2013sva,AlGendy:2014eua,Bernuzzi:2014kca,Takami:2014zpa}. These I-Love-Q relations were shortly after extended and shown to hold for NSs with a wider range of EoSs~\cite{lattimer-lim}, large and dynamical tidal deformations~\cite{maselli} and moderate magnetic fields~\cite{I-Love-Q-B}. The relation between the NS moment of inertia and quadrupole moment shows a clear spin-dependence~\cite{doneva-rapid,Pappas:2013naa,Chakrabarti:2013tca}, but the approximate EoS-universality is preserved even for rapidly-rotating NSs with a fixed dimensionless spin parameter~\cite{Pappas:2013naa,Chakrabarti:2013tca}. Approximately  universal relations were also shown to hold among higher multipole moments~\cite{Pappas:2013naa,Yagi:2014bxa} and in theories other than general relativity (GR), such as dynamical Chern-Simons gravity~\cite{I-Love-Q-Science,I-Love-Q-PRD}, Eddington-inspired Born-Infeld gravity~\cite{Sham:2013cya} and scalar-tensor theories~\cite{Pani:2014jra}.

The existence of such approximately universal relations was not anticipated. The EoS of a NS connects the microscopic physics governing its interior to macroscopic, measurable quantities like the mass and the radius. Not surprisingly, these macroscopic quantities depend sensitively on the specific NS EoS. For example, the mass and radius of a NS is constrained to lie on a particular mass-radius curve, which in turn depends sensitively on the EoS. In spite of this, it turns out that some quantities are affected by the EoS \emph{in roughly the same way} such that their interrelation is approximately EoS-independent. 

In an effort to understand this conclusion, Ref.~\cite{Stein:2013ofa} studied the multipole moments of rotating \emph{Newtonian stars} - stars constructed in the nonrelativistic limit, as a leading-order expansion in small compactness (ratio of stellar mass to stellar radius). By multipole moments, we mean the coefficients of a multipolar expansion of the exterior metric tensor or gravitational field in the weak- or far-field limit. Reference~\cite{Stein:2013ofa} found semianalytical relations between any arbitrary multipole moment and the first three moments: the mass, the spin and the quadrupole moment. The relations are semianalytic because they depend on solutions to the \emph{Lane-Emden} equation, which can in general be obtained only numerically, except for very few polytropic EoSs. Knowledge of the first three multipole moments of a NS allows for the semianalytic, full determination of all higher moments for a given EoS. 

The precise functional dependence of the $\ell^{th}$ moment on the mass, the spin, and the quadrupole moment was expected to depend on the NS EoS. Reference~\cite{Stein:2013ofa} showed that if one models NSs with polytropic EoSs, i.e.~EoSs of the form $p = K \rho^{1 + 1/n}$, with $p$ and $\rho$ the fluid's internal pressure and rest mass density, $K$ a constant and $n$ the \emph{polytropic index}, then the relations are very weakly dependent on $n$. In fact, Ref.~\cite{Stein:2013ofa} showed that for a range of polytropic indices consistent with that expected for realistic NS EoSs, $n \in [0.4,0.9]$, knowledge of the first three moments allows for the determination of the next $7$ with $\sim 5\%$ accuracy. This implies that we can {\emph effectively} infer the gravitational properties of NSs, regardless of our EoS ignorance, if we can we measure its first three moments.

We here continue the study of NSs in the nonrelativistic limit, since this is a \emph{controlled and accurate} approximation for the calculations carried out in this paper. In this limit, we expand all expressions in $C \ll 1$, where $C$ is the NS compactness. For NSs, $C = {\cal{O}}(10\%)$ and thus this is a good approximation. The approximation is also ``controlled'' because, in principle, higher-order in $C$ corrections could be calculated. Moreover, Ref.~\cite{Yagi:2014bxa} showed that relativistic corrections become less important when considering higher multipole moments. This is because the latter depend more sensitively on the outer regions inside the star, where the gravitational field is weaker and relativistic corrections less important~\cite{Yagi:2014qua}.

The main goal of this paper is to extend the analysis of Ref.~\cite{Stein:2013ofa} in two ways: 
\begin{enumerate}
\item[(i)] to increase the realism of the treatment by relaxing certain assumptions on the form of the EoS, and 
\item[(ii)] to increase the practicality of the result by providing a fully analytic treatment. 
\end{enumerate}
The former is achieved by studying whether the approximately universal relations of Ref.~\cite{Stein:2013ofa} continue to hold if one considers realistic EoSs. Reference~\cite{Stein:2013ofa} modeled NSs through an EoS characterized by a \emph{single} polytropic index. Instead, we will here use the parametrization of~\cite{Read:2008iy}, where the EoS is modeled through a set of piecewise polytropes. This parametrization has been shown to reproduce a variety of numerically-calculated, realistic NS EoSs to within a few percent accuracy.

We find that the approximate universality in the relations between multipole moments persists for realistic EoSs; our results match the ones obtained in~\cite{Stein:2013ofa} to a good accuracy. This is because the nonrelativistic approximation forces the NS central density to be small, which implies that the EoS becomes \emph{effectively} well-approximated by a single polytrope. Furthermore, the polytropes that contribute the most at low central densities are characterized by similar polytropic indices of $n\sim0.5$, for most EoSs. It should be emphasized that these results are \emph{not} a consequence of the nonrelativistic limit alone~\cite{Yagi:2014bxa}.

The second goal consists of deriving an approximate, but purely analytic set of relations among multipole moments. We model NSs with a single polytropic EoS and analytically perturb the Lane-Emden equation about a fiducial $n=0$ polytrope, where the background equations can be solved analytically. We then solve the perturbed equation to obtain fully analytical, approximately universal relations and compare them to the semianalytic relations of~\cite{Stein:2013ofa}, given in terms of Lane-Emden solutions.

We find that the purely analytic relations can reproduce the semianalytic ones obtained through the Lane-Emden solution to an accuracy of ${\cal{O}}(1\%)$ or better in the entire range of polytropic indices of interest for NS studies ($n\in[0.4,0.9]$). Furthermore, we find that the linear-order perturbation vanishes in the relation between the moment of inertia and quadrupole moment and in the slow-rotation limit, which provides further analytic understanding of the approximate universality.

The remainder of this paper presents the details of these calculations and it is organized as follows.
In Sec.~\ref{moments}, we explain the realistic extension of~\cite{Stein:2013ofa} and present the multipole moments relations for realistic EoS. 
In Sec.~\ref{perturbation}, we describe the practical extension of~\cite{Stein:2013ofa} and present the perturbative expansion of the single polytrope EoS. 
In Sec.~\ref{conclusions}, we conclude and discuss possible avenues for future work.
Throughout the paper, we use geometric units where $G=1=c$.

\section{Approximately Universal Relations among Multipole Moments with a Realistic EoS}
\label{moments}

We extend Ref.~\cite{Stein:2013ofa} by studying the approximately universal relations among multipole moments for Newtonian stars with realistic NS EoSs. We first define the Newtonian mass multipole moments and the Newtonian analogue of current multipole moments, and explain the Newtonian 3-hair relations found in Ref.~\cite{Stein:2013ofa}. We then introduce a piecewise polytropic EoS that can accurately reproduce various realistic NS EoSs~\cite{Read:2008iy} and extend the three-hair relation of~\cite{Stein:2013ofa} for piecewise polytropes. Having derived the relation between the stellar spin angular velocity and ellipticity, we present the results and study the approximate EoS-universality. We also look at the spin-dependence of the relations and compare our results with those reported in previous literature.

\subsection{Newtonian multipole moments and the three-hair relations}

We consider the nonrelativistic approximation where quantities are expanded in powers of the NS compactness $C = M_\nr/R_\nr \ll 1$, where $M_\nr$ is the mass and $R_\nr$ is the radius of the nonrotating NS. In this approximation, the metric tensor, or equivalently the gravitational field, in the exterior of a NS can be multipolarly expanded. The constant coefficients in this expansion are the (exterior) multipole moments of the spacetime, or simply the Newtonian moments when working to leading order in the non relativistic approximation.

Assuming uniform rotation, the Newtonian mass moments are defined through~\cite{Stein:2013ofa}
\be
M_{\ell} = 2 \pi \int^{\pi}_{0} \int^{R_{*}(\theta)}_0 \! \! \rho(r,\theta) \;  P_{\ell}(\cos{\theta}) \sin{\theta} \; r^{\ell + 2} \; d\theta dr,
 \label{Ml-integral}
\ee
and the Newtonian analogue of current moments are~\cite{Stein:2013ofa}
\be
S_{\ell} = \frac{4 \pi}{\ell+1} \Omega \int^{\pi}_{0} \int^{R_{*}(\theta)}_0 \rho(r,\theta) \frac{ d P_{\ell}(\cos{\theta})}{d \cos{\theta}} \sin^{3}{\theta} r^{\ell + 3} d\theta dr,
\label{Sl-integral}
\ee
where $R_{*}(\theta)$ is the surface of the star, $\rho(r,\theta)$ is the mass density, $\Omega$ is the surface spin angular velocity and $P_{\ell}(\cos{\theta})$ are Legendre polynomials. Following~\cite{Lai:1993ve}, we assume that the isodensity surfaces inside the star are self-similar ellipsoids and separate the integrals through the coordinate transformation
\be
r = \tilde{r} \Theta(\cos{\theta}) = \tilde{r} \sqrt{\frac{1-e^2}{1-e^2(1-\cos^2{\theta})}},
\ee
where $e$ is the star's ellipticity, which satisfies $R=a_1 (1-e^2)^{1/6}$ where $R$ is the geometric mean radius and $a_1$ is the semimajor axis. Transforming to this coordinate system and following the calculation of~\cite{Stein:2013ofa}, we find
\be
M_{2\ell + 2} = (-1)^{\ell + 1} \frac{4 \pi}{2 \ell + 3} e^{2 \ell +2} \sqrt{1 - e^2} R_{2 \ell + 2}\label{Mofell},
\ee
and
\be
S_{2\ell + 1} = (-1)^{\ell} \frac{8 \pi}{2 \ell + 3} \Omega(e) e^{2 \ell} \sqrt{1 - e^2} R_{2 \ell + 2}\label{Sofell},
\ee
where we have defined
\be
R_{\ell} = \int^{a_1}_{0} \rho(\tilde{r}) \tilde{r}^{\ell + 2} d \tilde{r},
\ee
and we recall that $\Omega(e)$ is the spin angular velocity of the star. The only quantity that depends on the EoS here is $R_{\ell}$, while Eqs.~\eqref{Mofell} and~\eqref{Sofell} are valid for any EoS.

We can now modify the approximately universal relations obtained in~\cite{Stein:2013ofa} to arbitrary EoSs. Defining the dimensionless multipole moments as
\be
\bar{M}_{\ell} = (-1)^{\ell/2} \frac{M_{\ell}}{M_0^{\ell + 1} \chi^{\ell}}, \quad \bar{S}_{\ell} = (-1)^{\frac{\ell -1}{2}} \frac{S_{\ell}}{M_0^{\ell + 1} \chi^{\ell}},\label{bardefs}
\ee
with $\chi \equiv S_1/M_0^2$, we can easily verify that
\be
\bar{M}_{2\ell + 2} = \bar{M}_2 \bar{S}_{2\ell + 1},
\ee
as first derived in~\cite{Stein:2013ofa}. Furthermore, we can show that
\be
\label{eq:2nd-univ}
\bar{M}_{2\ell + 2} = \bar{A}_{\ell} \; ( \bar{S}_{2\ell + 1})^{1+1/\ell},
\ee
where $\bar{A}_{\ell}$ is an EoS-dependent quantity given by
\be
\bar{A}_{\ell} = \frac{1}{3}\left( \frac{2\ell + 3}{3}\right)^{1/\ell} \frac{R_{2}^{1+1/\ell} R_{2\ell + 2}^{-1/\ell}}{R_0}.\label{abardef}
\ee
We have verified that for an EoS with a single polytropic index the above quantity reduces to Eq.~(17) of~\cite{Stein:2013ofa}. 

With the new definition of $\bar{A}_{\ell}$, the derivation of three-hair relations analogous to the ones obtained in~\cite{Stein:2013ofa} is straightforward:
\be
\bar{M}_{2\ell + 2} + i \bar{S}_{2\ell + 1} = \bar{B}_{\ell} \bar{M}^{\ell}_{2} (\bar{M}_{2} + i\bar{S}_{1})\label{hair-relation},
\ee
or alternatively
\be
M_{\ell} + i \frac{q}{a}S_{\ell} = \bar{B}_{\lfloor \frac{\ell-1}{2}\rfloor} M_0 (iq)^{\ell},
\ee
where $\bar{B}_{\ell} \equiv (\bar{A}_{\ell})^{-\ell}$, $a \equiv S_1/M_0$, $iq \equiv \sqrt{M_2/M_0}$, and $\lfloor \rfloor $ is the floor operator.

In order to evaluate $\bar{A}_{\ell}$ and quantify the degree of EoS independence in the no-hair relations between multipoles, we need to calculate the matter density of the star as a function of its radius. In the nonrelativistic, Newtonian limit, the internal structure of a star can be described by the Newtonian version of the \emph{Tolman-Oppenheimer-Volkoff} (TOV) equation 
\be
\frac{dp}{dr} = -\frac{\rho(r) m(r)}{r^2}\label{tov-def}\,,
\ee
which is nothing but the equation for hydrostatic equilibrium. In the above equation, $p(r)$ is the pressure, $\rho(r)$ the mass density and $m(r)$ is the mass inside a sphere with a radius $r$. Physically, this equation implies that the pressure gradient must balance the gravitational force exactly, in order for the star to be in equilibrium. This equation is underdetermined and cannot be solved unless we also specify the EoS of the star $p(\rho)$. 

\subsection{Parametrization of realistic EoSs}

Realistic EoSs can be parametrized by piecewise polytropes as shown in~\cite{Read:2008iy}, and we will use this parametrization to compute Eq.~\eqref{abardef} for a number of different EoSs. A generic piecewise polytropic representation of an EoS is simply
\be
p(\rho)=K_i \rho^{\Gamma_i}, \qquad \rho_{i-1} \leq \rho \leq \rho _i\,,
\ee
where the polytropic index $n_i$ is related to the adiabatic index $\Gamma_i$ through $n_i = 1/(\Gamma_i - 1)$. Reference~\cite{Read:2008iy} found that this piecewise polytropic EoS is an accurate representation of tabulated, realistic EoSs if one uses three pieces, $i \in \{1,2,3\}$, connected at fixed densities $\rho_1 = 10^{14.7} \rm{g/cm}^3$ and $\rho_2 = 10^{15.0} \rm{g/cm}^3$. Continuity of the pressure fixes the values of the constants $K_{i}$. 

Any given EoS is then fully determined by the choice of  $\{p_1, \Gamma_1, \Gamma_2, \Gamma_3\}$, where $p_1 \equiv p(\rho_1)$ and multiple choices that accurately fit tabulated data are given in Table III of~\cite{Read:2008iy}. We here use the subset of EoSs in Table III of~\cite{Read:2008iy} that give a maximum NS mass above $2M_{\odot}$. This is reasonable, given the recent observation of a pulsar with approximately that mass~\cite{2.01NS}. Table~\ref{eos} summarizes these EoSs and the $4$ parameters that describe them. A separate EoS is used for the crust region of the NS, but this will not contribute significantly to the no-hair relations, since there is barely any matter density in that region, relative to the interior of the star.  
\begin{table}
\begin{centering}
\begin{tabular}{ccccccc}
\hline
\hline
\noalign{\smallskip}
 EoS   &&&  $\log{p_1}$ &  $\Gamma_1$ & $\Gamma_2$ & $\Gamma_3$  \\
\hline
\noalign{\smallskip}
SLy~\cite{SLy} &&& 34.384 & 3.005 & 2.988 & 2.851\\
AP3~\cite{APR} &&& 34.392 & 3.166 & 3.573 & 3.281\\
AP4~\cite{APR} &&& 34.269 & 2.830 & 3.445 & 3.348\\
WFF1~\cite{Wiringa:1988tp} &&& 34.031 & 2.519 & 3.791 & 3.660\\
WFF2~\cite{Wiringa:1988tp} &&& 34.233 & 2.888 & 3.475 & 3.517\\
ENG~\cite{1996ApJ...469..794E} &&& 34.437 & 3.514 & 3.130 & 3.168\\
MPA1~\cite{1987PhLB..199..469M} &&& 34.495 & 3.446 & 3.572 & 2.887\\
MS1~\cite{Mueller:1996pm} &&& 34.858 & 3.224 & 3.033 & 1.325\\
MS1b~\cite{Mueller:1996pm} &&& 34.855 & 3.456 & 3.011 & 1.425\\
H4~\cite{Lackey:2005tk} &&& 34.669 & 2.909 & 2.246 & 2.144\\
ALF2~\cite{Alford:2004pf} &&& 34.616 & 4.070 & 2.411 & 1.890\\
\noalign{\smallskip}
\hline
\hline
\end{tabular}
\end{centering}
\caption{Piecewise polytropic fit for the EoSs used here~\cite{Read:2008iy}. $p_1$ is in units of $\rm{dyn}/\rm{cm}{}^2$.}
\label{eos}
\end{table}
%

\subsection{Newtonian mass density \\ with realistic EoSs}

We now use the piecewise polytropic representation of the EoSs to find how 
the density and pressure behave as a function of radius, i.e.~the Lane-Emden equation. 
Writing the mass density as 
\be
\rho = \rho_i \theta_i^{n_i},
\ee
where $\rho_i$ is the transition density of the $i$th piece of the piecewise EoS, we insert this equation into Eq.~\eqref{tov-def} and arrive at 
\be
\frac{1}{\xi^2}\frac{d}{d\xi}\left(\xi^2\frac{d\theta_i}{d\xi}\right)+\theta_i^{n_i}=0,\label{laneemden-def}
\ee
where we have made the change of variables $r=\alpha_i\xi$, with
\be
\alpha_i^2=\frac{K_i(n_i+1)\rho_i^{\frac{1}{n_i}-1}}{4\pi}.
\ee
Equation~\eqref{laneemden-def} is the \emph{Lane-Emden} equation with initial conditions $\theta_i(0)=1$, $\theta_i'(0)=0$ and its solution gives the density and the pressure of a star described by an EoS of polytropic index $n_i$ as a function of its radius~\cite{shapiro-teukolsky}. With the parametrization of the EoSs used in this paper, solving for $\rho(r)$ requires solving the Lane-Emden equation for the $3$ different adiabatic indices corresponding to each piece of the EoS, and stitching the solutions at the appropriate values of the density by imposing the continuity and differentiability conditions on the physical density at each transition radius. 

In order to solve the Lane-Emden equation and calculate $\rho(r)$, we need to specify the central density of the star. What central density should we select to make a meaningful comparison between stars with different EoSs? In Ref.~\cite{Stein:2013ofa} the authors were able to circumvent this problem by noting that for a single polytrope, the central density cancels out from Eq.~\eqref{abardef}. In our case, however, the value of $\rho_c$ determines the relative importance on the $3$ polytropes, and, thus, affects our results. Equation~\eqref{hair-relation} and the fact that $\bar{S}_1=1$, however, makes it clear that the comparison should be carried out \emph{with central densities that lead to the same value of $\bar{M}_{2}$}. Using Eqs.~\eqref{Mofell}-\eqref{Sofell} and the first of Eqs.~\eqref{bardefs}, we can write the dimensionless mass quadrupole as
\be
\bar{M}_2=\frac{3}{4}\frac{e^2}{\Omega(e)^2} \frac{R_0}{R_2},
\ee
where the dependence of the ellipticity on the spin angular velocity of the star is encoded in $\Omega(e)$, and it is EoS-dependent. As we consider different values of the set $\{p_{1},\Gamma_{1},\Gamma_{2},\Gamma_{3}\}$ that approximate different tabulated EoSs, we will ensure that the central density is chosen so as to keep $\bar{M}_{2}$ the same from EoS to EoS.  

\subsection{Calculation of $\Omega(e)$}
\label{omega}

From the previous subsection, it is clear that if we wish to find the central density that leads to the same value of the dimensionless quadrupole moment as we vary the EoS we need an expression for the angular frequency as a function of ellipticity for a generic EoS. Lai et al.~\cite{Lai:1993ve} calculate this relation for a single polytrope by finding the star configuration that minimizes the energy. Here we follow their derivation and generalize their result to arbitrary EoSs.

To leading Newtonian order in a nonrelativistic expansion, the energy of a rotating star with a given mass and spin angular momentum can be approximated by
\be
\label{eq:EofUWT}
E=U+W+T,
\ee
where $U$ is the internal energy of the thermal motion of the particles, $W$ is the gravitational potential energy, and $T$ is the rotational energy.

Once the total energy $E$ has been expressed in terms of the ellipticity, the energy minimization requirement~\cite{Lai:1993ve} gives
\be
\frac{\pd E}{\pd e}=\frac{\pd E}{\pd \lambda}=0,\label{energy-condition}
\ee 
where we follow the notation of~\cite{Lai:1993ve} and define $\lambda \equiv (1-e^2)^{1/3}$. The calculation is simplified slightly, by noting that
\be
U=\int u\, dm, \quad du=\frac{p(r)}{\rho(r)^2}d\rho.
\ee
Since $U$ is ellipticity-independent, it can be neglected. 

The next term in Eq.~\eqref{eq:EofUWT} is the gravitational potential energy $W$. Lai et al.~\cite{Lai:1993ve} show that the gravitational energy of a deformed object is given by $W_{\nr} (R) g(e)$, where $W_{\nr}$ is the gravitational energy of the nonrotating configuration, and $g(e)$ is a function of the ellipticity 
\be
g(e) \equiv \frac{\sin^{-1} e}{e} \left(1-e^2 \right)^{1/6}.
\ee
The non rotating gravitational potential energy is given by
\begin{align}
W_{\nr} (R_\nr) &= -\int_0^{M_\nr} \frac{m}{r} dm = -12 \pi \int_0^{R_\nr} p(r) \; r^2 dr, \nn \\
\label{grav-energy} 
\end{align}
where we have used the Newtonian TOV equation [Eq.~\eqref{tov-def}], the mass continuity equation, and in the last step, we integrated by parts and used that $p=0$ at the surface of the star.

The last contribution to the energy we need to consider is the rotational energy, given by~\cite{Lai:1993ve}
\be
T = \frac{S_1^2}{2I} = \lambda \frac{S_1^2}{2 I_s},\label{rot-energy} 
\ee
where $I$ is the moment of inertia, and $I_s$ is the moment of inertia of a spherical star with the same volume, given by
\ba
I_s=\frac{2}{3} \int r^2 dm &=& \frac{8 \pi}{3} (1-e^2)^{5/6} R_2,\label{moment-inertia}
\ea
where we have used the coordinate transformation $r = (1-e^2)^{1/6} \tilde{r}$. Note that $R_2$ depends on $a_1$ and, thus, on the ellipticity. But the quantity $(1-e^2)^{5/6} R_2$ is ellipticity-independent, and can be neglected when differentiating with respect to $e$ or $\lambda$. Adding Eqs.~\eqref{grav-energy} and~\eqref{rot-energy} and imposing the minimum energy condition in Eq.~\eqref{energy-condition} we get for the spin angular velocity
\be
\label{eq:Omega-e}
\Omega(e)^2 = \frac{9}{4} \frac{f(e)^2}{(1-e^2)^{5/6} R_2} \; \int_0^{R} p(r) \; r^2 dr, 
\ee
where $f(e)$ is given by~\cite{Stein:2013ofa}
\begin{align}
f(e) ={}& \left[ - 6 e^{-2} \left(1 - e^{2}\right)
\right.
\nn \\
& \left.
{}+2 e^{-3} \left(1 - e^{2}\right)^{1/2} \left(3 - 2 e^{2}\right) \arcsin{(e)} \right]^{1/2}.
\end{align}
%
\subsection{Approximately universal relations}
\label{relations}

Having devised a way to calculate $\bar{M}_2$ given a central density and spin angular momentum, we can now calculate the EoS-dependent coefficient of the $3$-hair relations in Eq.~\eqref{hair-relation}. We choose to work in the slow-rotation (and thus the small-ellipticity) limit, an approximation that will be justified later. In this limit, $\bar{M}_2$ reduces to
\be
\label{M2bar-slow}
\bar{M}_2=\frac{5}{8} R_0 \left(\int^{R_{\nr}} p(r) \, r^2 dr\right)^{-1}\,.
\ee
In what follows we select a value for $\bar{M}_2$ and, for each EoS, we find the value of the central density that leads to stars of the same $\bar{M}_2$ value.

\subsubsection{$\bar{M}_2=10$ case}
\label{m2bar_10}

Since we are working in the Newtonian limit, we expect our results to agree with the fully relativistic results of~\cite{Yagi:2014bxa} in this limit. From Fig.~1 of \cite{Yagi:2014bxa}, stars with $\bar{M}_2=10$ are in the nonrelativistic Newtonian regime, and the analysis of this paper should be accurate. 
\begin{figure}[t]
\begin{center}
\includegraphics[width=\columnwidth,clip=true]{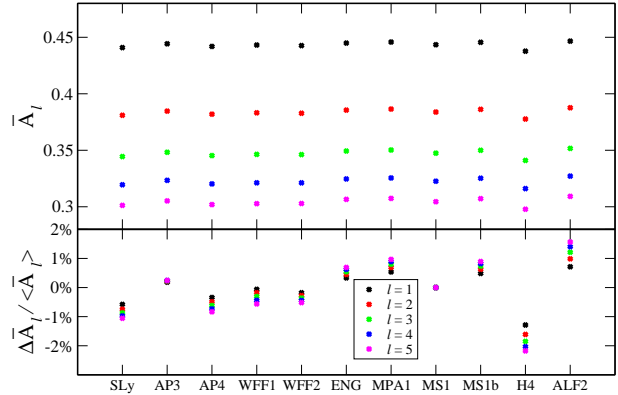}
\caption{\label{fig:Abar_M2bar_10} (Top panel) Coefficient $\bar{A}_{\ell}$ for the various EoSs in Table~\ref{eos} for stars with $\bar{M}_2=10$ in the slow rotation limit. (Bottom panel) Relative fractional difference between the values of $\bar{A}_{\ell}$ from the top panel and the median of all EoSs for a given ${\ell}$ denoted as $\langle\bar{A}_{\ell}\rangle$. The fractional error never exceeds $\sim2\%$, in agreement with the results of~\cite{Stein:2013ofa}.}
\end{center}
\end{figure}

The top panel of Fig.~\ref{fig:Abar_M2bar_10} shows the value of $\bar{A}_{\ell}$, for the various EoSs considered here, while the bottom panel shows the fractional error relative to the median of $\bar{A}_{\ell}$ for all EoSs with the same value of $\ell$. The fractional error is less than $\sim 2 \%$, as first obtained in~\cite{Stein:2013ofa}.

The fact that the approximate universality holds to such good accuracy should be no surprise, given the relatively high value of $\bar{M}_2$. For most EoSs, this high value of $\bar{M}_2$ is achieved by choosing central densities below $10^{15}$g/cm$^3$, the dividing density between the second and the third polytrope. Therefore, for such low central densities, most EoSs are effectively parametrized by only two polytropes. Table~\ref{eos} shows that most EoSs considered here have an average polytropic index in this region $(n_1+n_2)/2 \sim 0.5$. The only exception is H4 with $(n_1+n_2)/2= 0.66$. Not surprisingly, this EoS has the largest fractional error.

\subsubsection{Case $\bar{M}_2=100$}
\label{m2bar_100}

\begin{figure}[t]
\begin{center}
\includegraphics[width=\columnwidth,clip=true]{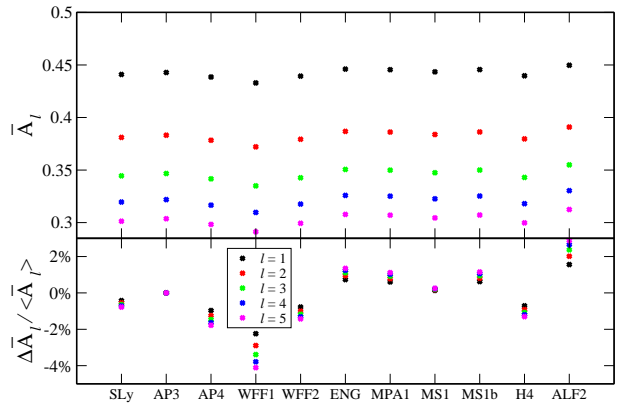}
\caption{\label{fig:Abar_M2bar_100}
Same as Fig.~\ref{fig:Abar_M2bar_10} but with $\bar{M}_2=100$. The maximum fractional error is about $(3-4)\%$, in agreement with the results of~\cite{Stein:2013ofa} for a single polytrope.}
\end{center}
\end{figure}

NSs with $\bar{M}_2=100$ can be captured very well by the Newtonian description we employ here. Compared to the $\bar{M}_2=10$ case, all stars have even lower central densities, thus bringing all EoSs into the density regime where they are all very well described by a single polytrope. The top panel of Fig.~\ref{fig:Abar_M2bar_100} shows $\bar{A}_{\ell}$ for stars of $\bar{M}_2=100$, while the bottom panel shows its fractional error relative to the EoSs median. Again, as expected from the discussion of the previous subsection, the WFF1 $(n_1=0.66)$ and ALF2 ($n_1=0.33)$ EoSs present the largest fractional errors.
%

\subsubsection{Case $\bar{M}_2=1$}
\label{m2bar_1}

NSs with $\bar{M}_2=1$ have large compactnesses, and we do not expect them to be adequately described by the leading order Newtonian terms in a non relativistic expansion. However, it is still interesting to investigate the Newtonian relations in such a high central density regime, where all $3$ polytropes contribute. 

\begin{figure}[t]
\begin{center}
\includegraphics[width=\columnwidth,clip=true]{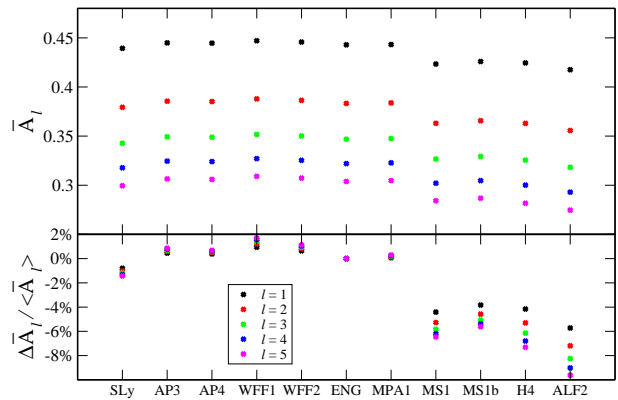}
\caption{\label{fig:Abar_M2bar_1}  
Same as Fig.~\ref{fig:Abar_M2bar_10} but with $\bar{M}_2=1$. For the majority of the EoSs the fractional error is in the range $(1-2)\%$, a consequence of the fact that most polytropes have indices $n\sim0.5$. However, the last four EoSs have a fractional error of $\sim 5\%$. This is due to the large central density of these stars and the large indices of the $3^{\rm{rd}}$ (innermost) polytrope.}
\end{center}
\end{figure}

Figure~\ref{fig:Abar_M2bar_1} studies this case, with the top panel showing the values of $\bar{A}_{\ell}$, and the bottom panel giving its fractional error relative to the median of all EoSs. Clearly, the approximate universality of $\bar{A}_{\ell}$ is preserved to a very good approximation, with most EoSs having a fractional error of $\sim 2\%$. However, the MS1, MS1b, H4 and ALF2 EoSs have larger fractional errors, $\sim (4-8)\%$. 

This is due to two effects: (i) NSs with these EoSs and $\bar{M}_2=1$ have central densities much higher than the dividing density of the second and the third polytropes $10^{15}$g/cm$^3$, which means that it is mostly the $3^{\rm{rd}}$ polytrope that contributes to $\bar{A}_{\ell}$, and (ii) the $3^{\rm{rd}}$ polytropic indices are $n_3 = 3.07$ for MS1, $n_3 = 2.35$ for MS1b, $n_3 = 0.87$ for H4, and $n_3= 1.12$ for ALF2; these values are very different from the $n\sim 0.5$ effective index of all other EoSs, leading to quantitatively different results.

%

\subsection{Effect of spin}
\label{spin}

The analysis presented previously assumed that the spin angular velocity of the star is small relative to its mass. In this limit, the ellipticity cancels out from $\bar{M}_2$ [Eq.~\eqref{M2bar-slow}]. Here, we show that this assumption is well justified, since the neglect of spin effects introduces an error in $\bar{A}_{\ell}$ that is smaller than that caused by the EoS uncertainty. 
\begin{figure}[t]
\begin{center}
\includegraphics[width=\columnwidth,clip=true]{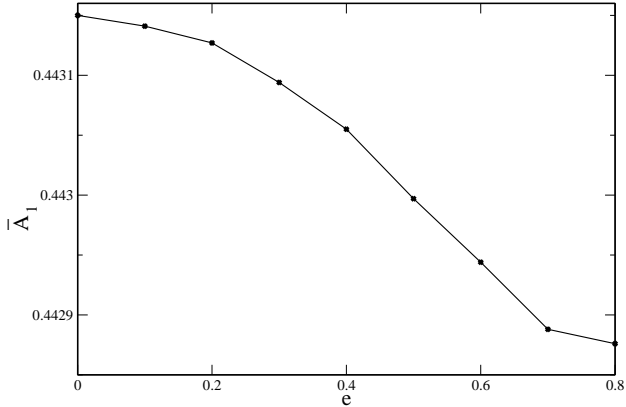}
\caption{\label{fig:Abar_ellipticity_M2bar_10_EoS_3}  Coefficient $\bar{A}_1$ as a function of the ellipticity of the star. This quantity is insensitive to rotational corrections.}
\end{center}
\end{figure}

Figure~\ref{fig:Abar_ellipticity_M2bar_10_EoS_3} shows the coefficient $\bar{A}_1$ for a NS with the WFF1 EoS as a function of ellipticity calculated from Eq.~\eqref{eq:Omega-e} without imposing the slow-rotation approximation. Even at extremely high values of ellipticity, and large spin angular velocities, the spin correction to the slowly-rotating $\bar{A}_1$ result does not exceed $\sim 0.05\%$, which is $2$ orders of magnitude smaller that the fractional error due to the EoS uncertainty. Such results are consistent with previous results in the literature; Ref.~\cite{Stein:2013ofa} found that the multipole relations are completely spin-independent for Newtonian single polytropes in the elliptical isodensity approximation, while Refs.~\cite{Pappas:2013naa,Yagi:2014bxa} showed that such relations up to the hexadecapole order $(\ell =4)$ are spin-insensitive for NSs in full GR.

\section{Perturbative Analysis about a Background Polytrope}
\label{perturbation}

In this section, we describe a practical extension of Ref.~\cite{Stein:2013ofa} by looking for a set of approximate but purely analytic relations among Newtonian multipole moments with a single polytropic EoS. We will achieve this goal by perturbing the Lane-Emden equation about a fiducial $n=0$ polytrope and solving the perturbed equation analytically. We will compare the analytic relations in terms of the perturbed Lane-Emden solution with the semianalytic ones in terms of the numerical Lane-Emden solution found in~\cite{Stein:2013ofa} for various polytropic indices.

Restricting ourselves to a single polytropic EoS of the form $p = K \rho^{1+1/n}$, Eq.~\eqref{abardef} simplifies to 
\be
\bar{A}_\ell \to \bar{A}_{n,\ell} =
\frac{  (2 \ell +3)^{1/\ell}}{3^{(1+1/\ell)}}
\frac{{\cal{R}}_{n,2}^{1+1/\ell}{\cal{R}}_{n,2+2 \ell}^{-1/\ell}}{{|\vartheta'(\xi_{1})| \; \xi_{1}^{2}}}\,,
\ee
where the radial integral is defined by
\be
{\cal{R}}_{n,\ell} =  \int_{0}^{\xi_{1}} 
 \left[\vartheta(\xi)\right]^{n} \xi^{\ell+2} d \xi.
\ee

For an $n=0$ polytrope, we can solve the Lane Emden equation analytically. Using the boundary conditions $\vartheta(0)=1$ and $\vartheta'(0)=0$, we find $\vartheta^{(n=0)}=1-\xi^2/6$ with $\xi_1^{(n=0)}=\sqrt{6}$. From these, we can calculate $\bar{A}_{0,\ell}$ analytically to find
\begin{align}
\label{A0sol}
\bar{A}_{0,\ell} ={}& \frac{1}{5} \left[\frac{(2 \ell + 3) (2 \ell + 5)}{15}\right]^{1/\ell}.
\end{align}

\begin{figure*}[htb]
\centering
\includegraphics[width=\columnwidth,clip=true]{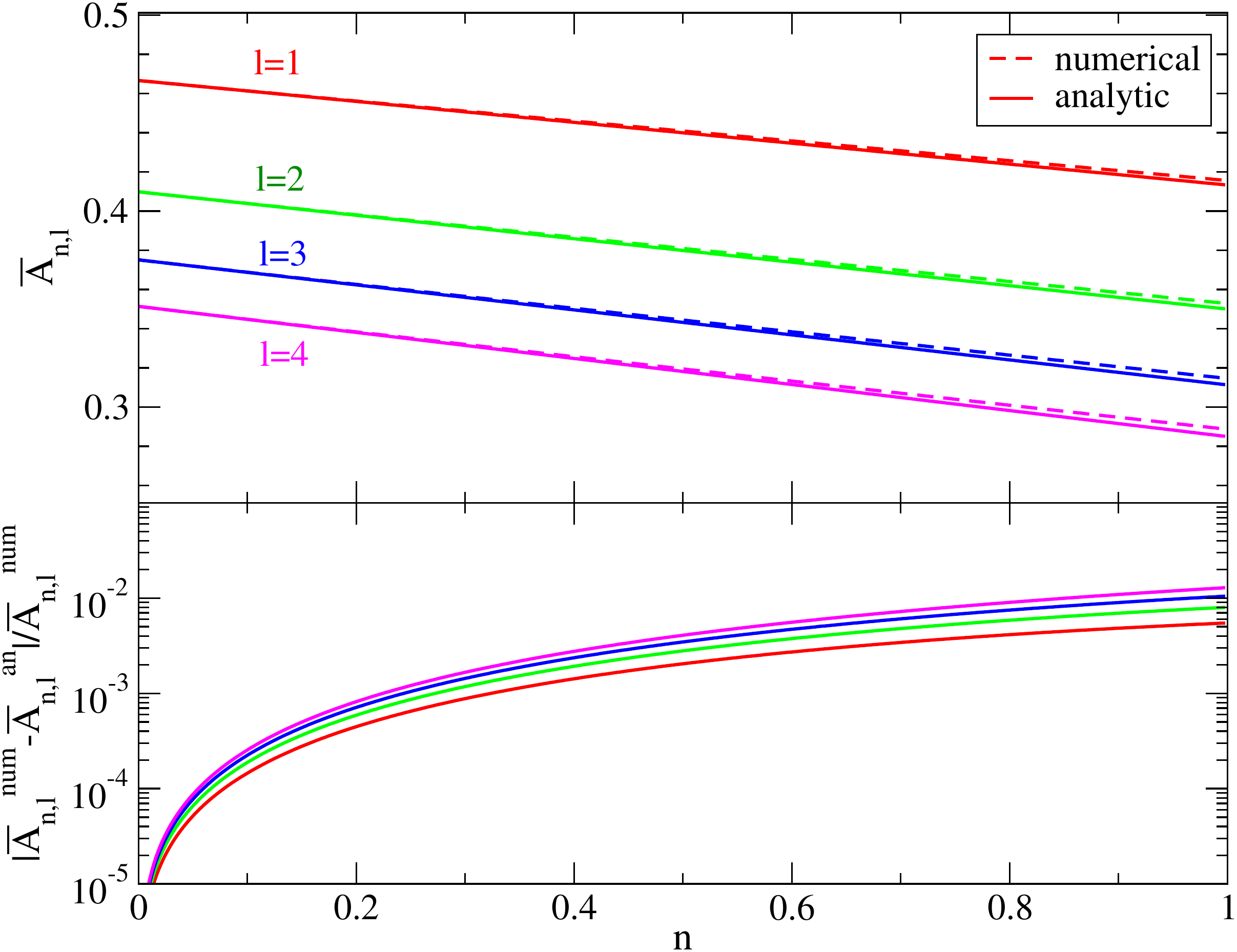}
\includegraphics[width=\columnwidth,clip=true]{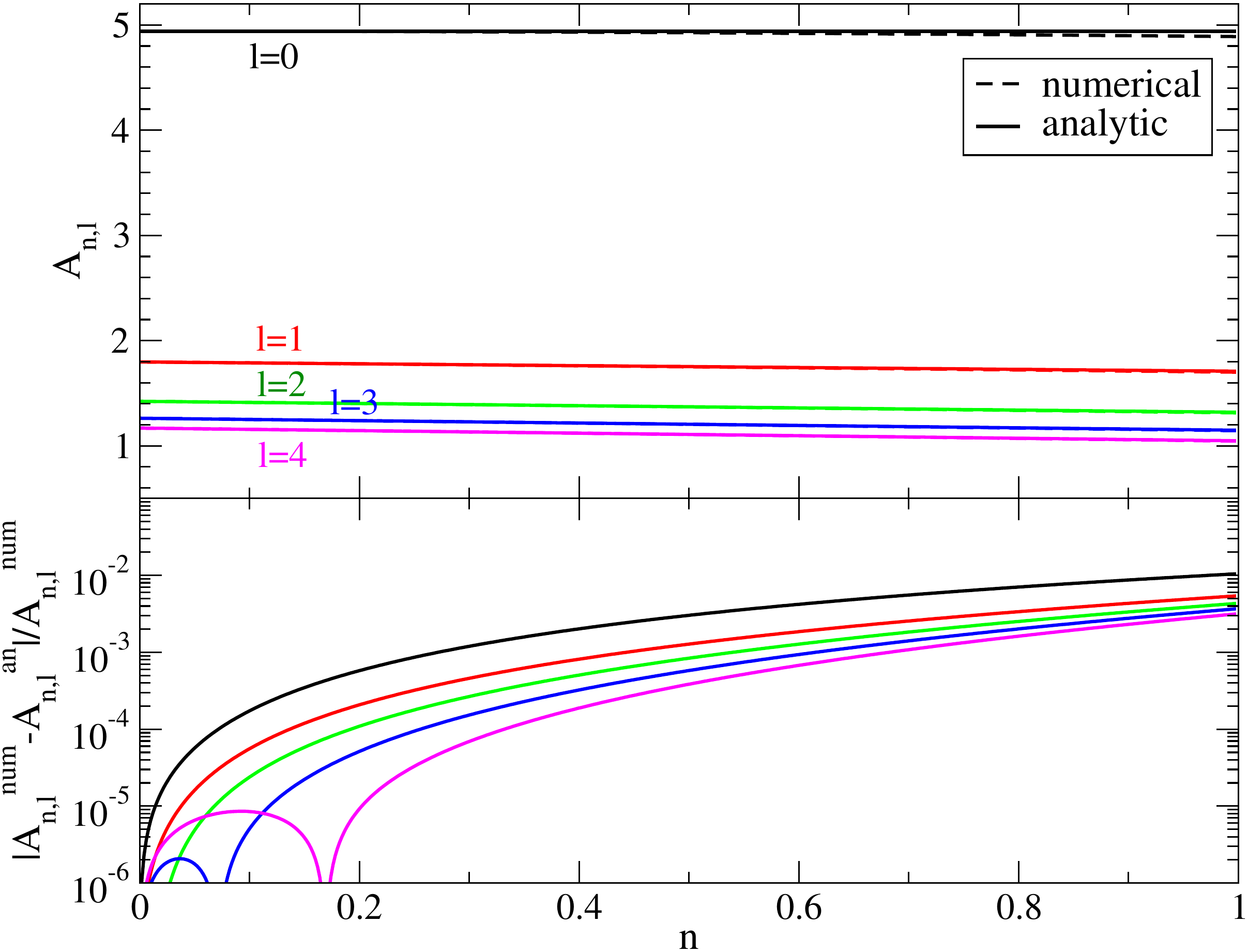}
\caption{
(Top Left) $\bar{A}_{n,\ell}$ vs $n$ obtained numerically found in~\cite{Stein:2013ofa} (dashed) and analytically about $n=0$ (solid) for various $\ell$. The analytic result obtained here matches beautifully with the numerical one, even at $n = 1$. (Bottom Left) Fractional relative difference between the analytic and numerical values of $\bar{A}_{n,\ell}$. Such difference is $\mathcal{O}(1\%)$ at most for all $\ell \leq 4$ at $n=1$. (Right) Same as the left panel but for $A_{n,\ell}$.
\label{fig:Abar-comparison-numerical-analytic-n0}
}
\end{figure*}

We now carry out a perturbative analysis purely analytically in the nonrelativistic limit by assuming a density profile of the form 
\be
\rho = \rho_{c} \vartheta^{n},
\ee
with 
\be
\label{eq:n-pert}
n=\tilde{n} + \epsilon_2\,, \quad \vartheta = \tilde{\vartheta} + \epsilon_2 \delta \vartheta + \mathcal{O}(\epsilon_2^2),
\ee
where $\tilde{n}$ is a background polytropic index, $\tilde\vartheta$ is a background solution to the Lane-Emden equation, $\epsilon_2$ is a perturbation to the polytropic index and $\delta \vartheta$ is a perturbation to the Lane-Emden solution. Inserting Eq.~\eqref{eq:n-pert} into Eq.~\eqref{laneemden-def} with $\vartheta_i \to \vartheta$ and $n_i \to n$ and expanding to $\mathcal{O}(\epsilon_2)$, we find~\cite{1978SvA....22..711S,Bender:1989xe,Seidov:2004tt} 
\be
2 \tilde\vartheta \frac{d\delta\vartheta}{d\xi} + \tilde\vartheta \xi \frac{d^2 \delta\vartheta}{d\xi^2} + \tilde n \tilde\vartheta^{\tilde n} \xi \delta\vartheta + \ln(\tilde\vartheta) \tilde\vartheta^{\tilde n+1} \xi = 0.
\ee
Since $\tilde{\vartheta}$ already satisfies the initial conditions of the Lane-Emden solution, we require that $\delta \vartheta(0) = 0 = \delta \vartheta'(0)$. 

We consider the $\tilde{n}=0$ case as the background polytrope, where $\bar{A}_{0,\ell}$ was already presented in Eq.~\eqref{A0sol}. The perturbed Lane-Emden equation can be solved to find~\cite{1978SvA....22..711S}
\ba
\delta \vartheta^{(\tilde n=0)} &=& - \frac{\xi^2}{18} \left[ 3 \ln \left( 1-\frac{\xi^2}{6} \right) - 5 \right] +3 \ln \left( 1-\frac{\xi^2}{6} \right)  \nn \\
& & -4 + 4 \sqrt{6} \frac{\tanh^{-1} \left( \xi/\sqrt{6} \right)}{\xi}.
\ea
The perturbation to the NS surface is $\xi_1 = \tilde{\xi}_1 + \epsilon_2 \delta \xi_1 + \mathcal{O}(\epsilon_2^2)$, where $\tilde{\xi}_{1} = \sqrt{6}$, while $\delta \xi_{1}$ can be derived by requiring that $\theta(\xi_{1}) = 0$; we find
\be
\delta \xi_1 = - \frac{\delta\vartheta(\tilde\xi_1)}{\tilde\vartheta'(\tilde\xi_1)},
\ee
which for an $\tilde{n}=0$ background yields~\cite{1978SvA....22..711S}
\ba
\delta \xi_1^{(\tilde n=0)} &=& 2\sqrt{6} \ln 2 - \frac{7}{\sqrt{6}}.
\ea
One has to be careful when evaluating $\delta\vartheta(\tilde\xi_1)$ at an $\tilde n=0$ background, because it naively diverges, although the limit is perfectly well defined. 

With these corrections at hand, we can now evaluate the corrections to the second approximately universal relation in Eq.~\eqref{eq:2nd-univ}. Expanding $\bar{A}_{0,\ell} = \tilde{\bar{A}}_{0,\ell} + \epsilon_2 \delta \bar{A}_{0,\ell} + \mathcal{O}(\epsilon_2^2)$ we find
\begin{align}
\frac{\delta \bar{A}_{0,\ell}}{\tilde{\bar{A}}_{0,\ell}} &=
\frac{1}{15 \ell (5+2 \ell) (3+8 \ell + 4 \ell^2)} 
\left[ \left(225 + 690 \ell + 540 \ell^2 
\right. \right. \nn \\
&\left. + 120 \ell^3) \; H \! ( \ell- 1/2 ) 
+ (450 + 1380 \ell + 1080 \ell^2 \right. 
\nn \\ 
& \left. 
+ 240 \ell^3) \ln 2 - 1126 \ell - 1572 \ell^2 - 584 \ell^3 - 48 \ell^4\right],
\label{ratio}
\end{align}
where $H(\ell) \equiv \sum_{k=0}^{\ell} 1/k$ is the $\ell$th harmonic number, and $\tilde{\bar{A}}_{0,\ell}$ is the unperturbed $\bar{A}_{0,\ell}$ coefficient evaluated at the background $\tilde n=0$ polytrope. For fractional arguments, the harmonic number can be evaluated by analytic continuation through its standard integral representation, $H(1/2) = 2 - 2 \ln{2}$.  One may naively think that Eq.~\eqref{ratio} diverges for $\ell = 0$ because the second term is proportional to $2 \ln{2}/\ell$. Careful inspection, however, reveals that the first term asymptotes to $H(-1/2)/\ell$, which cancels this divergence exactly and yields a finite result. 

The top left panel of Fig.~\ref{fig:Abar-comparison-numerical-analytic-n0} shows $\bar{A}_{n,\ell}$ versus $n$ obtained in~\cite{Stein:2013ofa} for various values of $\ell$ by solving the Lane-Emden equation numerically, together with $\tilde{\bar{A}}_{0,\ell} + n \delta \bar{A}_{0,\ell}$ obtained analytically above.  The analytic perturbative result matches the numerical one to a very good approximation. The bottom left panel of Fig.~\ref{fig:Abar-comparison-numerical-analytic-n0} shows the fractional difference between analytic and numerical values of $\bar{A}_{n,\ell}$ for various $\ell$ modes. The difference between the analytic and the numerical results increases as one increases $n$ and $\ell$.  However, even at $n=1$, the fractional difference between numerical results and our analytical approximations is at most $\mathcal{O}(1\%)$ up to $\ell =4$. Such perturbed relations about $n=0$ are in fact good enough to represent the full numerical result, at least up to $\ell \leq 4$, because the ellipsoidal isodensity approximation is itself valid only up to $\mathcal{O}(1\%)$~\cite{Lai:1993ve}. One can perform a similar analysis to obtain an analytic relation perturbed about the $n=1$ polytrope, but the result is more complicated and not more illuminating, so we do not include it here.

Within a slow-rotation approximation, we can derive a similar approximately universal relation, related to the I-Love-Q ones~\cite{Stein:2013ofa} 
\be
\label{eq:2nd-no-hair-new}
|M_{2 \ell + 2}| =
A_{n,\ell} \left[ \frac{|S_{2\ell+1}|^{5(\ell+1)}}{M_0^{\ell+4} (M_0 \Omega)^{\ell+1}} \right]^{\frac{1}{5\ell+2}} \left[ 1+{\cal{O}}(M_0 \Omega)^{2} \right],
\ee
where the dimensionless coefficient $A_{n,\ell}$ is given by
\be
\label{eq:A-def}
A_{n,\ell} =
\left\{ \!\!\left[\frac{25\left(5 - n\right)^2}{1152} \right]^{\ell + 1}\!\!\!\!\!\!\!\!
\frac{(2 \ell + 3)^{3} \left({\cal{R}}_{n,2}\right)^{2 \ell + 2}}{\left({\cal{R}}_{n,2\ell+2}\right)^{3} \xi_{1}^{2 \ell -4}|\vartheta'(\xi_{1})|^{2 \ell -1}}  
\right\}^{\frac{1}{5 \ell + 2}} \!\!\!\!\!\!\!\!\,.
\ee
The $\ell=0$ case directly corresponds to the $\bar{I}$--$\bar{M}_2$ relation.
 We perform a similar perturbative analysis as that done on $\bar{A}_{n,\ell}$ about a fiducial polytropic index $\tilde{n}$. Perturbing $A_{n,\ell}$ as $A_{n,\ell} = \tilde A_{n,\ell} + \delta A_{n,\ell}$, we find
\be
\frac{\delta A_{0,\ell}}{\tilde{A}_{0,\ell}} = \frac{3 \ell}{5 \ell +2} \frac{\delta \bar{A}_{0,\ell}}{\tilde{\bar{A}}_{0,\ell}}.
\ee
In particular, $\delta A_{0,0}/\tilde{A}_{0,0}=0$. This result is not trivial and it is related to the fact that Eq.~\eqref{ratio} has a finite limit at $\ell =0$. One can in fact show that this is not true for $\ell > 0$, with $\delta A_{0,\ell}/\tilde{A}_{0,\ell}$ scaling as $\epsilon_2$.  This means that around $\tilde{n}=0$, $A_{n,0}$ scales at least quadratically in $n$ (not linearly). 

The right panel of Fig.~\ref{fig:Abar-comparison-numerical-analytic-n0} shows the numerical values of $A_{n,\ell}$, together with the analytic ones about $n=0$ found above. As in the $\bar{A}_{n,\ell}$ case, the analytic result reproduces the numerical results to $\mathcal{O}(1\%)$ accuracy. Unlike in the $\bar{A}_{n,\ell}$ case, however, the fractional difference for $A_{n,\ell}$ decreases as we increase $\ell$. This suggests that the fractional difference for $A_{n,\ell}$ is of $\mathcal{O}(1\%)$ at most.

\section{Conclusions} 
\label{conclusions}

In this paper, we extended the analysis of Ref.~\cite{Stein:2013ofa} by considering realistic NS EoSs. In particular, instead of considering a single polytropic EoS, we here considered piecewise polytropic EoSs found to accurately reproduce tabulated NS EoS~\cite{Read:2008iy}. With such EoSs, we showed that the relations between multipole moments of a Newtonian rigidly rotating NS and the first three (mass, spin and quadrupole moment) are still approximately EoS-independent. In particular, we found that knowledge of the mass, the spin and the quadrupole moment of a NS would allow us to calculate the next $7$ moments to within $2\%$ accuracy or better. Moreover, we confirmed that the multipole relations are spin-insensitive, which is consistent with the results~\cite{Stein:2013ofa,Pappas:2013naa,Yagi:2014bxa}.

We also extended the results of Ref.~\cite{Stein:2013ofa} by deriving practical, purely analytic relations among multipole moments for Newtonian polytropes. In particular, we perturbed the Lane-Emden equation about an $n=0$ polytrope and used the perturbed analytic solution to derive purely analytic 3-hair relations for Newtonian polytropes. We compared these results with those in~\cite{Stein:2013ofa}, obtained by solving the Lane-Emden equation numerically, and found that the new analytic relations reproduce the numerical ones to $\sim 1\%$ accuracy. Moreover, we found that the linear perturbation to the relation between the moment of inertia and quadrupole moment in the slow-rotation limit vanishes, which provides further analytic evidence and a deeper understanding of the approximate EoS-universality.

The results presented here are part of the recent effort to find approximately universal relations between NS observables~\cite{I-Love-Q-Science,Stein:2013ofa}. Such relations can simplify the analysis of astrophysical data considerably, leading to a faster and more accurate determination of the EoS through the elimination of degeneracies, as discussed in~\cite{I-Love-Q-Science,I-Love-Q-PRD,baubock,Stein:2013ofa,Psaltis:2013fha}. 

Future work could concentrate on improving the analytic relations by extending the perturbation to the next order~\cite{Seidov:2004tt}. One can also consider a perturbation about an $n=1$ polytrope~\cite{1978SvA....22..711S,Bender:1989xe} and combine the result with that of a perturbation about an $n=0$ polytrope presented here. Since we here worked in the Newtonian limit, one could also compute relativistic corrections to our results. These corrections would be proportional to the NS compactness and they can be obtained through the use of the relativistic integral representation of exterior multipole moments found in~\cite{Ryan:1996nk}. 

Recently, Ref.~\cite{Yagi:2014qua} proved that the self-similarity of isodensity surfaces inside a NS plays a crucial role in the approximate universality. Therefore, it would be interesting to perform a similar Newtonian analysis to that in~\cite{Stein:2013ofa} and this paper but for differentially-rotating stars. Such rotation naturally breaks the self-similarity of the isodensity contours and one would thus expect the universality to be lost. One could also study the relations for differentially-rotating NSs in full GR and compare them against Newtonian relations. Such a goal can be achieved by adopting the $j$-constant rotation law~\cite{komatsu_eh1989} and constructing a differentially-rotating NS solution in full GR~\cite{1989MNRAS.239..153K,Stavridis:2007xz,Passamonti:2007td}.

Finally, another possible avenue for future work is to perform a similar Newtonian analysis of other types of universal relations, such as those in NS oscillation modes~\cite{andersson-kokkotas-1998,kokkotas-living,tsui-leung,lau}. Reference~\cite{Yagi:2013sva} reported universal relations among various tidal deformability parameters for NSs. Such relations should be given semianalytically in the Newtonian limit in terms of the solution to the Clairaut-Radau equation~\cite{1978trs..book.....T,brooker-olle,mora-will}. It would be interesting to study such relations with realistic NS EoSs, and perturb the equation around an $n=0$ polytrope to obtain purely analytic relations, just like the extended multipole relations in the Newtonian limit found here.

\acknowledgments
We would like to thank Leo Stein and George Pappas for comments on this manuscript. We would also like to thank Charles Kankelborg for comments on our numerical implementation. K.C. acknowledges support from the Onassis Foundation. N.Y. acknowledges support from NSF grant PHY-1114374, NSF CAREER Grant PHY-1250636 and NASA grant NNX11AI49G. 

\bibliography{review}
\end{document}